\documentclass[number,longtitle,sort&compress]{elsarticle}

\usepackage{tikz}
\usetikzlibrary{patterns}
\usetikzlibrary{plotmarks}
\usepackage{graphicx}
\usepackage{hyperref}
\usepackage{subfig}
\usepackage{mciteplus}
\usepackage{varwidth}
\usepackage{float}
\usepackage{lineno}
\usepackage{amsmath}
\usepackage{amssymb}

\newcommand{\Cerenkov}{\mbox{\v{C}erenkov}}
\newcommand{\degree}{\mbox{$^\circ$}}

\graphicspath{{./figs/}}

\journal{Nuclear Instruments and Methods in Physics Research A}

\begin{document}

\begin{frontmatter}



\title{A Threshold Gas \Cerenkov{} Detector for the Spin Asymmetries of the Nucleon Experiment}

\author[temple]{Whitney R. Armstrong\corref{cor1} }
\ead{whit@temple.edu}

\author[seoul]{Seonho Choi}

\author[temple]{Ed Kaczanowicz}
\author[temple]{Alexander Lukhanin}

\author[temple]{Zein-Eddine Meziani}

\author[temple,jlab]{Brad Sawatzky}

\cortext[cor1]{Corresponding author}


\address[temple]{Department of Physics, Temple University, Philadelphia, PA, 19122, USA}
\address[seoul]{Seoul National University, Seoul 151-747, Korea}
\address[jlab]{Thomas Jefferson National Accelerator Facility, Newport News, 
Va, 23606, USA}



\begin{abstract}
We report on the design, construction, commissioning, and performance of a 
threshold gas \Cerenkov{} counter in an open configuration, which operates in a 
high luminosity environment and produces a high photo-electron yield. Part of a 
unique open geometry detector package known as the Big Electron Telescope 
Array, this \Cerenkov{} counter served to identify scattered electrons and 
reject produced pions in an inclusive scattering experiment  known as the Spin 
Asymmetries of the Nucleon Experiment  E07-003 at the Thomas Jefferson National 
Accelerator Facility (TJNAF) also known as Jefferson Lab.  The experiment 
consisted of a measurement of double spin asymmetries $A_{\parallel}$ and 
$A_{\perp}$ of a polarized  electron beam impinging on a polarized ammonia 
target. The \Cerenkov{} counter's performance is characterised by a yield of 
about 20 photoelectrons per electron or positron track. Thanks to this large 
number of photoelectrons per track, the \Cerenkov{} counter had enough 
resolution to identify electron-positron pairs from the conversion of photons 
resulting mainly from $\pi^0$ decays.  \end{abstract}

%

\begin{keyword}
Threshold Gas \Cerenkov{} Detector \sep SANE \sep Particle Identification
\PACS 29.40.Ka
\end{keyword}

\end{frontmatter}





\section{Introduction}

\begin{center}
\begin{figure}[h]
\centering
\includegraphics[width=\textwidth]{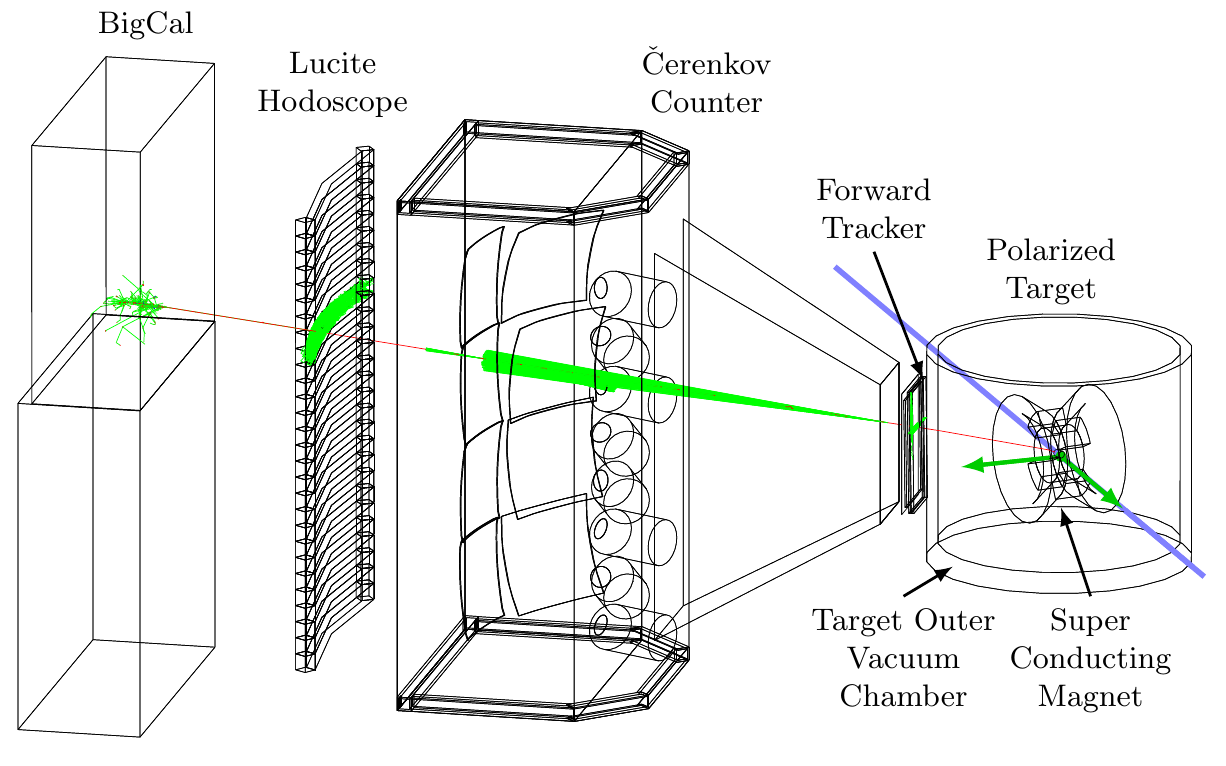}
\caption{ BETA detectors and polarized target.\label{fig:BETA0}}
\end{figure}
\end{center}

\Cerenkov{} counters have been used in high energy electron scattering 
experiments as part of detector stacks within shielded spectrometers, 
especially at low duty cycle beam facilities where the instantaneous beam 
current was at the milliamp level. With the advent of continuous electron beams 
with very high duty cycle it became feasible to use \Cerenkov{} counters in an 
open environment for the same average beam current. In the SANE experiment data 
were collected in Hall C at Jefferson Lab using a threshold gas \Cerenkov{} 
counter, as an integral part of a new detector package known as the Big 
Electron Telescope Array (BETA), which identified scattered electrons, measured
their scattered angle and their energy.  BETA, shown in figure~\ref{fig:BETA0}, 
consisted of a forward tracker, a gas \Cerenkov{} detector which is the focus 
of this report, a Lucite hodoscope and a lead-glass calorimeter.

Using a polarized electron beam and a polarized ammonia target,
The SANE collaboration~\cite{SANEproposal} set out to measure two observables,
known as double spin asymmetries, $A_{\parallel}$ and $A_{\perp}$. These are 
measured with two different configurations, parallel or perpendicular, of the 
spin directions of the beam and the target.  The electron beam's helicity was 
flipped while the spin direction of the target has been fixed. To detect the 
scattered electrons BETA was positioned at a central scattering angle of $40$ 
degrees, covering roughly $\pm5$\degree{} in scattering angle, and from about 
$700$ MeV up to about $2.5$ GeV in energy, thus covering a  kinematic range of  
Bjorken $x$ and four momentum transfer $Q^2$ corresponding to $0.3 < x < 0.8$  
and $2.5$~GeV$^2 < Q^2 < 6.5$~GeV$^2$. 

The gas \Cerenkov{} counter's role was to identify inclusively scattered 
electrons and to reject pions during the experiment, thus providing an 
efficient trigger for clean electrons in a high background environment.

A description of the SANE apparatus follows in section \ref{sec:BETA}. In 
addition to a mechanical overview, section \ref{sec:Design} discusses the 
design choices and their motivation.  Section~\ref{sec:construction} presents, 
in detail, the construction of the SANE \Cerenkov{} counter.  This is followed 
by a description of the detector's calibrations and initial commissioning.  
Electronics and calibrations are the focus of section~\ref{sec:PMTs}.  Due to 
differences in background rates, magnetic field, and particle trajectories, the 
performance during each  target field orientation is separately discussed in 
section~\ref{sec:Performance}. Our conclusions are described in 
section~\ref{sec:conclusion}.  

\section{SANE Apparatus}
\label{sec:BETA}

To reach the statistical precision required by SANE in a limited amount of 
incident beam time BETA, a device unique in its open configuration,  was built 
to provide the required angular and momentum acceptance for the experiment. 

\subsection{The BETA Detector Package}

\begin{center}
\begin{figure}[h]
\centering
\includegraphics[trim = 1mm 12mm  10mm 1mm,  clip, width=\textwidth]{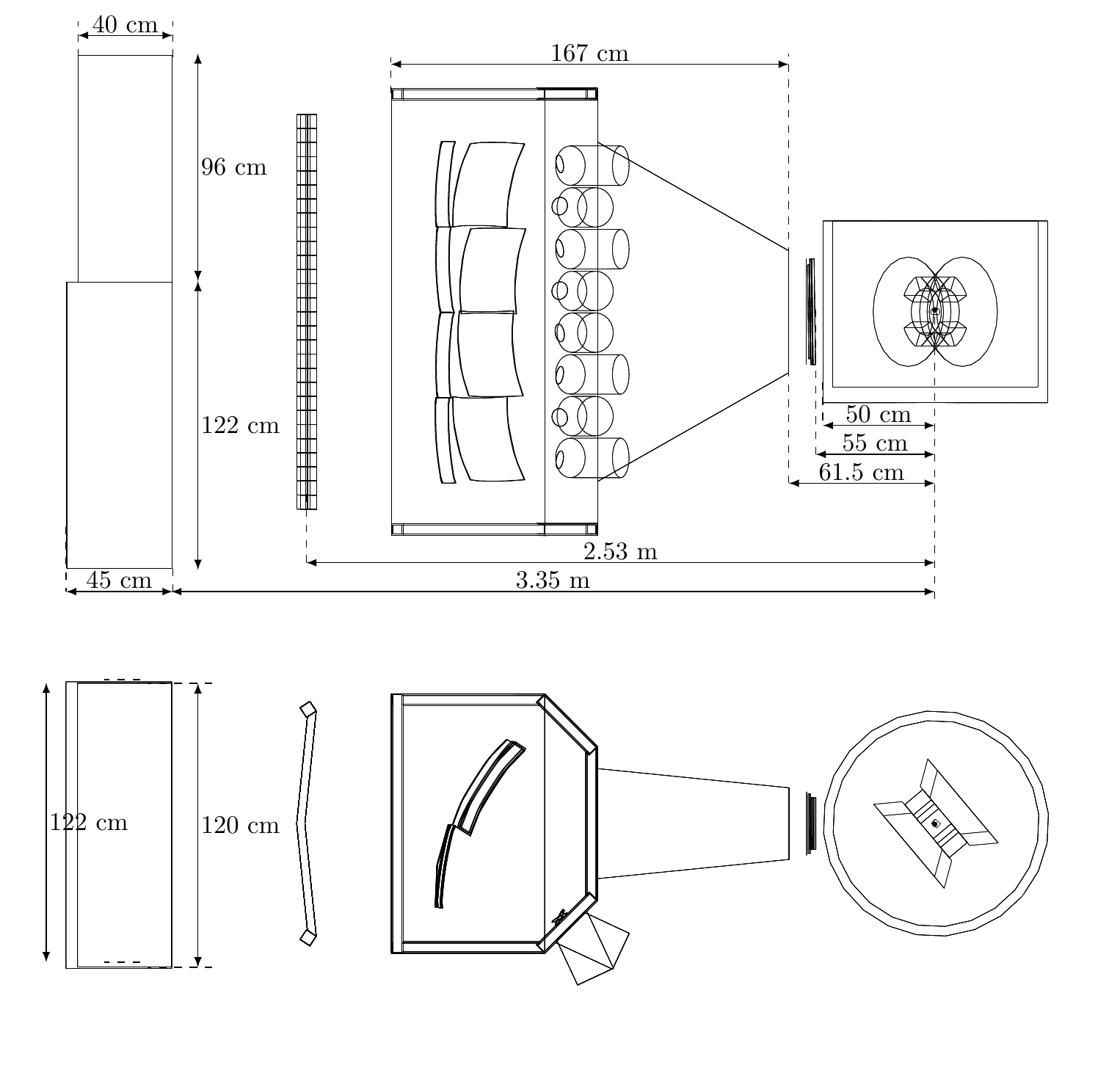}
\caption{BETA dimensions with side view (upper figure) and a top view (lower 
figure). Shown from left to right are the calorimeter, hodoscope, \Cerenkov{} counter, forward tracker and polarized target.
\label{fig:BETADimensions} }
\end{figure}
\end{center}

BETA comprises four detectors, a forward tracker placed as close to the target 
as possible, followed by a \Cerenkov{}  counter, a Lucite hodoscope and a large 
electromagnetic calorimeter dubbed BigCal.  As can be seen in 
Figure~\ref{fig:BETA0}, the front window of the \Cerenkov{} counter was 
positioned just behind the forward tracker while BigCal was set at roughly 
$3.5$m from the target.  The dimensions and positions of the detectors and 
target are shown in Figure~\ref{fig:BETADimensions}.  

BigCal provided  position and energy measurements of electrons and photons. It 
consisted of 1792 lead glass blocks and was divided into two sections.  The 
lower section makes use of $3.8$~cm$\times3.8$~cm$\times45$~cm blocks arranged 
in a $32\times32$ grid, while the upper section contains  
$4$~cm$\times4$~cm$\times40$~cm blocks arranged in a $30\times24$ grid.

The Lucite hodoscope includes 28 curved Lucite bars with light guides mounted 
to edges cut at 45\degree{}. Stacked vertically, each bar was 6 cm tall and 3.5 
cm thick and provided a vertical position measurement. Photomultipliers (PMTs) 
were connected at both ends of the bar, providing an additional  position  
determination in the horizontal direction by taking the time difference between 
two discriminated PMT signals.

The forward tracker used wavelength shifting fibers glued to BC-408 plastic 
scintillator to detect the scattered particles as close to the target as 
possible. Two layers  of $3$~mm$\times3$~mm$\times22$~cm scintillators, stacked 
vertically and offset by $1.5$ mm along with a layer of 
$3$~mm$\times3$~mm$\times40$~cm scintillators piled horizontally, provided a 
position measurement with a resolution sufficient to distinguish between 
electron and positron trajectories of momenta, below $1$~GeV/c.

\subsection{Polarized Target }

Used in previous deep-inelastic scattering experiments that measured also 
double spin asymmetries at 
SLAC~\cite{Averett:1999nz,Abe:1998wq,Anthony:2002hy,McNulty:2002yf} and at 
Jefferson Lab~\cite{Wesselmann:2006mw}, the polarized target system used a 5.1 
T magnet to polarize ammonia through the mechanism of dynamic nuclear 
polarization. In addition to polarizing the target material, the presence of 
this magnetic field reduces background radiation by trapping charged particles 
with a momentum less than about $180$~MeV.  As will be discussed in 
section~\ref{sec:transverse}, magnetic trapping also increased background 
during transverse polarization running. Due to depolarization effects, the 
nominal production beam current was kept roughly at about $100$~nA which in 
turns limited the experimental luminosity to about $\sim$1.7$\times$10$^{35}$ 
cm$^{-2}$ s$^{-1}$ when using the $3$~cm thick ammonia target.

The experiment required two target configurations, (anti-)parallel, with the 
magnetic field pointed along the incoming electron beam direction, and 
transverse, where mechanical constraints limited the target angle to be 80 
degrees with respect to the beam direction. In addition to rotating the magnet, 
additional upstream chicane magnets were used to compensate for the beam 
deflection during transverse running. Deflection of background particles in the 
downstream target field required a non-standard beamline incorporating a helium 
gas bag and extended exit beam pipe.

\section{Design of the \Cerenkov{} Counter}
\label{sec:Design}

The primary requirements for the SANE gas \Cerenkov{} counter were to identify 
electrons with high efficiency (greater than $90$\%) while maintaining a pion 
rejection factor better than 1000:1 with software cuts.  The open configuration 
of the experiment required a high $\pi^\pm$ momentum threshold to reject much 
of the low energy background. The \Cerenkov{} counter needed to cover a rather 
large solid angle of about 200 msr and thus was placed
as close as possible to the target.

Charged pions needed to be rejected for momenta up to $5$~GeV. Pions above 
this momentum threshold should be extremely rare with a 4.7 GeV or 5.9 GeV 
incident electron beam and would be removed in any case with our software 
electron selection cuts. Looking at the thresholds for commonly used gases 
shown in Table~\ref{tab:certhresholds}, N$_2$ gas was selected as it best meets 
this rejection requirement.

\begin{table}[h]
\begin{center}
\begin{tabular}{|l|l|l|l|}
\hline
\emph{Charged particle}& N$_2$ & CO$_2$  & C$_4$F$_{10}$  \\
\hline
$e^+$, $e^-$     & 21 MeV    & 17 MeV  & 7.8 MeV  \\
$\mu^+$, $\mu^-$ & 4.33 GeV  & 3.5 GeV & 1.6 GeV \\
$\pi^+$, $\pi^-$ & 5.75 GeV  & 4.6 GeV & 2.1 GeV \\
\hline
\end{tabular}
\caption{\Cerenkov{} thresholds for charge particles in a N$_2$ gas at 
atmospheric pressure compared with other common gases.  
\label{tab:certhresholds}}
\end{center}
\end{table}

Operating at atmospheric pressure helped simplify the mechanical design and 
minimize window thicknesses. The choice of a N$_2$ gas \Cerenkov{} for radiator 
was a trade-off between \Cerenkov{} photon-yield, material budget, undesired 
scintillation, and (ultraviolet) transparency. Although N$_2$ gas is known to 
scintillate, the low density and optical design minimized the impact of the 
isotropic scintillation background (Section~\ref{sec:Optics}).  At $20$\degree 
C, the index of refraction of N$_2$ gas is $n=1.000279$, yielding a threshold 
velocity, $\beta_{0}$ , for \Cerenkov{} light emission by charged particles of
\begin{displaymath}
\beta_{0} = \frac{1}{n} = 0.999721.
\end{displaymath}
Furthermore, N$_2$ is inexpensive in comparison to other gasses making a gas 
recovery system unnecessary.

The leading mechanical design constraint was the optics design for focusing the 
\Cerenkov{} light onto PMTs.  Additional constraints on the mechanical design 
included a high rate of background at small angles, access to any mirrors and 
PMTs, and operation in the fringe magnetic field from the target.

A symmetric mirror-PMT design could not be considered as a viable option given 
the high background rates and long experimental run times in the area closer to 
the beam line. In order to allow for easy maintenance of the PMTs and to 
minimize worker radiation exposure\cite{ALARA} all PMTs were set on the large 
scattering angle side of the detector. Furthermore, with the PMTs on one side 
(as seen in figure \ref{fig:BETADimensions}), background shielding was 
consolidated into two large walls.

This asymmetric design called for elliptical mirrors on the far side of the 
\Cerenkov{} tank instead of spherical mirrors. Although complicating mirror 
procurement, these mirrors provided a focused light profile necessary to 
operate with three-inch diameter PMTs. Five-inch diameter tubes were initially considered, 
however their increased size made it very difficult to shield magnetically, and 
shield from background radiation. The polarized target's $5.1$ Tesla 
superconducting magnet sat just over one meter away necessitating a significant 
amount of mu-metal shielding which housed the PMTs. The mu-metal was rolled 
into $3$ mm thick cylinders to mitigate the effects of a $100-200$ Gauss 
magnetic field around the three-inch PMTs.

\section{\Cerenkov{} Construction}
\label{sec:construction}

\subsection{Mirror Optics}
\label{sec:Optics}

After an extensive detailed optical ray-trace analysis which included among other things 
the effect of the target magnetic field on the scattered electron trajectories, 
the shapes of the mirrors were determined\cite{BETAMirrorDesign}.  The 
optimized result called for eight, roughly 40 cm by 40 cm glass mirrors 
arranged in two overlapping columns of four mirrors to cover the rather large 
acceptance shown in Fig~\ref{fig:BETADimensions}. Four spherical mirrors  cover 
one column at large scattering angles and four elliptical mirrors cover another 
column at small scattering angles.  For machining purposes, the elliptical 
mirror was approximated with a spindle torus without any effect on the 
performance \cite{MirrorBlanks}. The dimensions of the mirrors are listed in 
Table~\ref{tab:mirrorDimensions}. Light from each mirror is focussed onto 
individual 3-inch quartz window photomultipliers (Photonis XP4318B).


\begin{table}[h]
\begin{center}
\begin{tabular}{|l|l|}
\hline
Spherical Mirror &  \\
\hline
radius of curvature & $92.0$ cm \\ 
vertical size & $36.5$ cm \\ 
horizontal size & $35.5$ cm \\ 
\hline \hline
Toroidal Mirror & \\
\hline
minor circle (`tube') radius &  $85.8$ cm  \\
circle of revolution (`donut') radius &  $25.8$ cm  \\ 
vertical size & $36.5$ cm \\ 
horizontal size & $43.0$ cm \\ 
\hline
x semiaxis (ellip. equiv.)& $97.0$ cm \\
y and z semiaxes (ellip. equiv.)& $86.0$ cm \\
\hline
\end{tabular}
\caption{Dimensions for the spherical and toroidal mirrors. Also the torus-approximated elliptical surface's semiaxes for a mirror in the x-y plane are given.  }
\label{tab:mirrorDimensions}
\end{center}
\end{table}
\begin{figure}[h]
\centerline{\includegraphics[width=\textwidth,scale=1.0,angle=0.0]{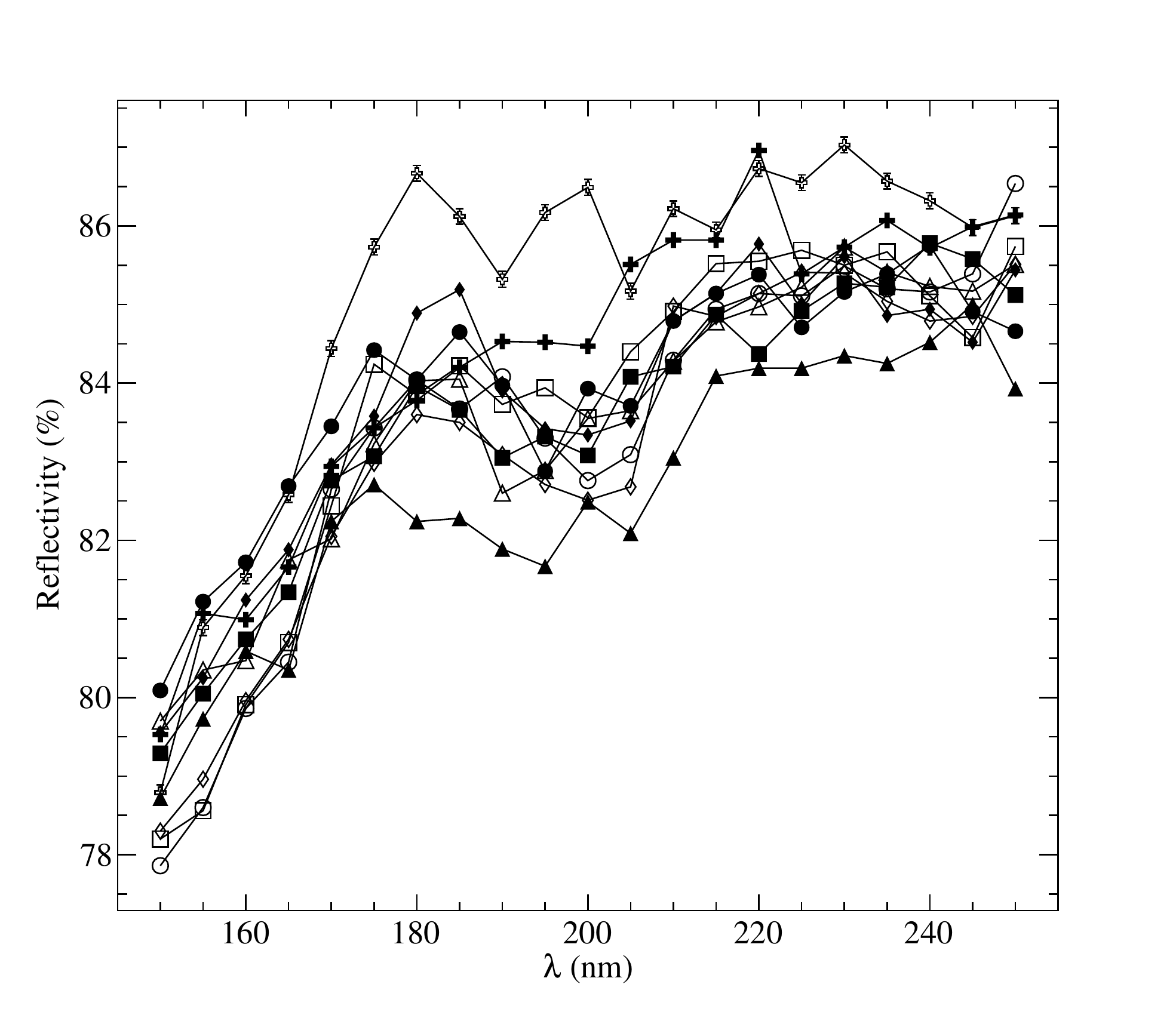}}
\caption{\label{fig:reflectivity}Mirror reflectivity as measured at CERN \cite{mirrorcoating}. }
\end{figure}

The glass mirrors were coated with aluminum and MgF$_2$ for maximum 
reflectivity into the far UV in order to compliment the sensitivity of the 
photomultiplier tube at these wavelengths. The reflectivity for each mirror was 
measured at CERN\cite{mirrorcoating} and shown in 
Figure~\ref{fig:reflectivity}.

The electrons of interest have momentum above 0.7~GeV/c and are deflected
by the target field less than a few degrees. Thus, to a good
approximation, the mirrors have been designed for point-to-point focusing
from the target cell to the photomultiplier photocathodes. This
permits the two towers of mirrors to be optimally aligned with a
small, bright light bulb located at the same target-mirror
distance. This geometry also permits good rejection of stray light
from scintillation and low energy $\delta$ rays (which are
preferentially emitted at angles several times larger than the
\Cerenkov{} cone).

\subsection{Mirror Alignment}

\begin{figure}[h]
\centering
\includegraphics[width=0.75\textwidth]{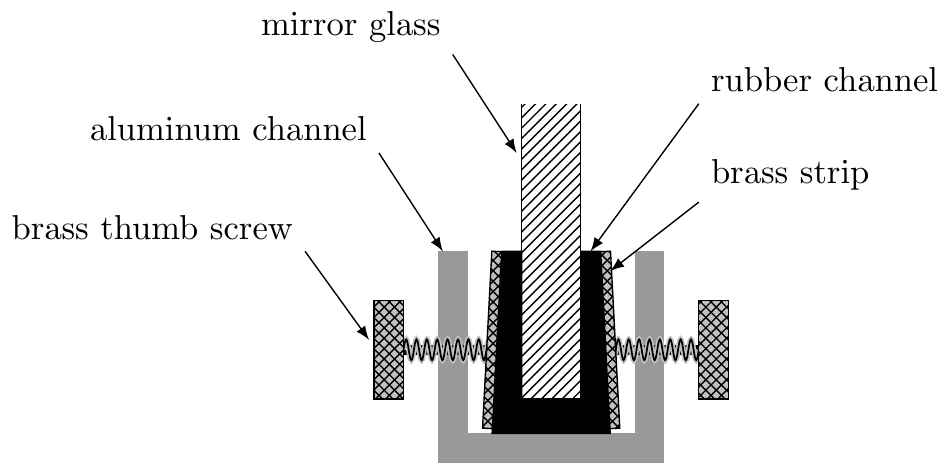}
\caption{A cross section view of the mirror holder u-channel. The channel was machined to match the mirror curvature. }
\label{fig:mirrorHolder}
\end{figure}
\begin{figure}[h]
 \centering
 \subfloat[Align top 
 mirrors.]{\includegraphics[width=0.3\textwidth]{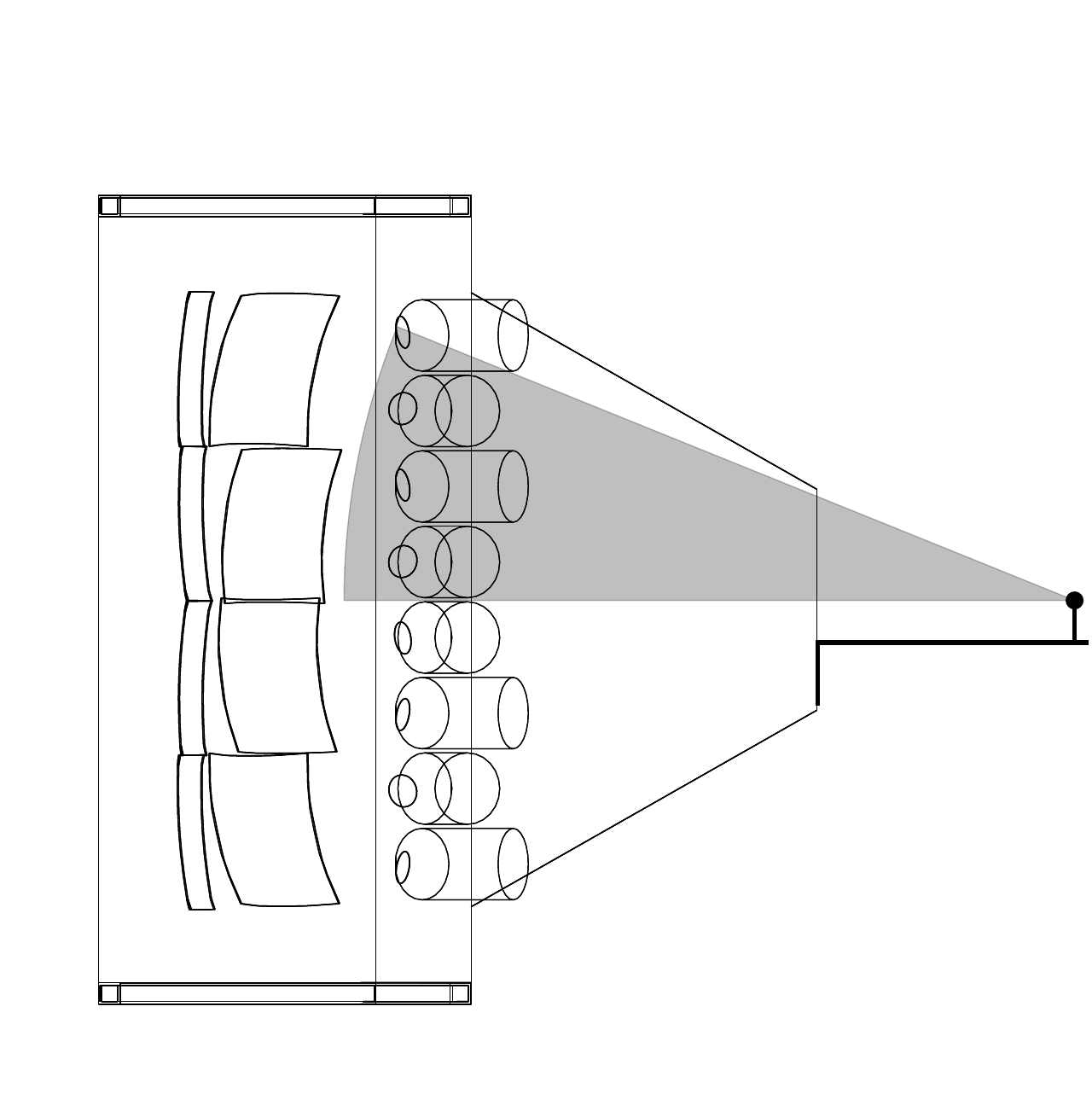} } 
 \subfloat[Align bottom mirrors.]{\includegraphics[width=0.3\textwidth]{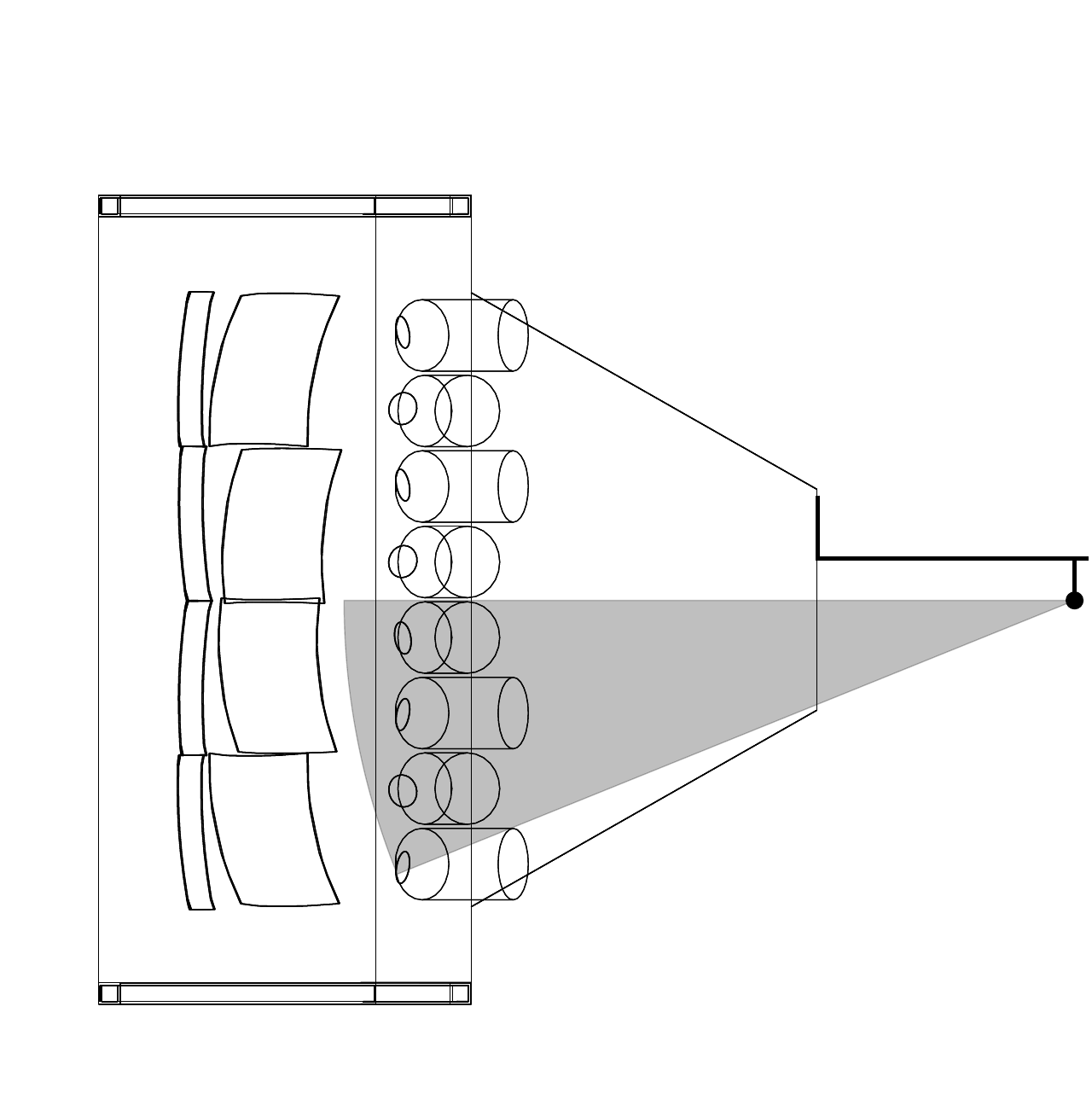} } 
 \subfloat[Completed mirror alignment.]{\includegraphics[width=0.35\textwidth]{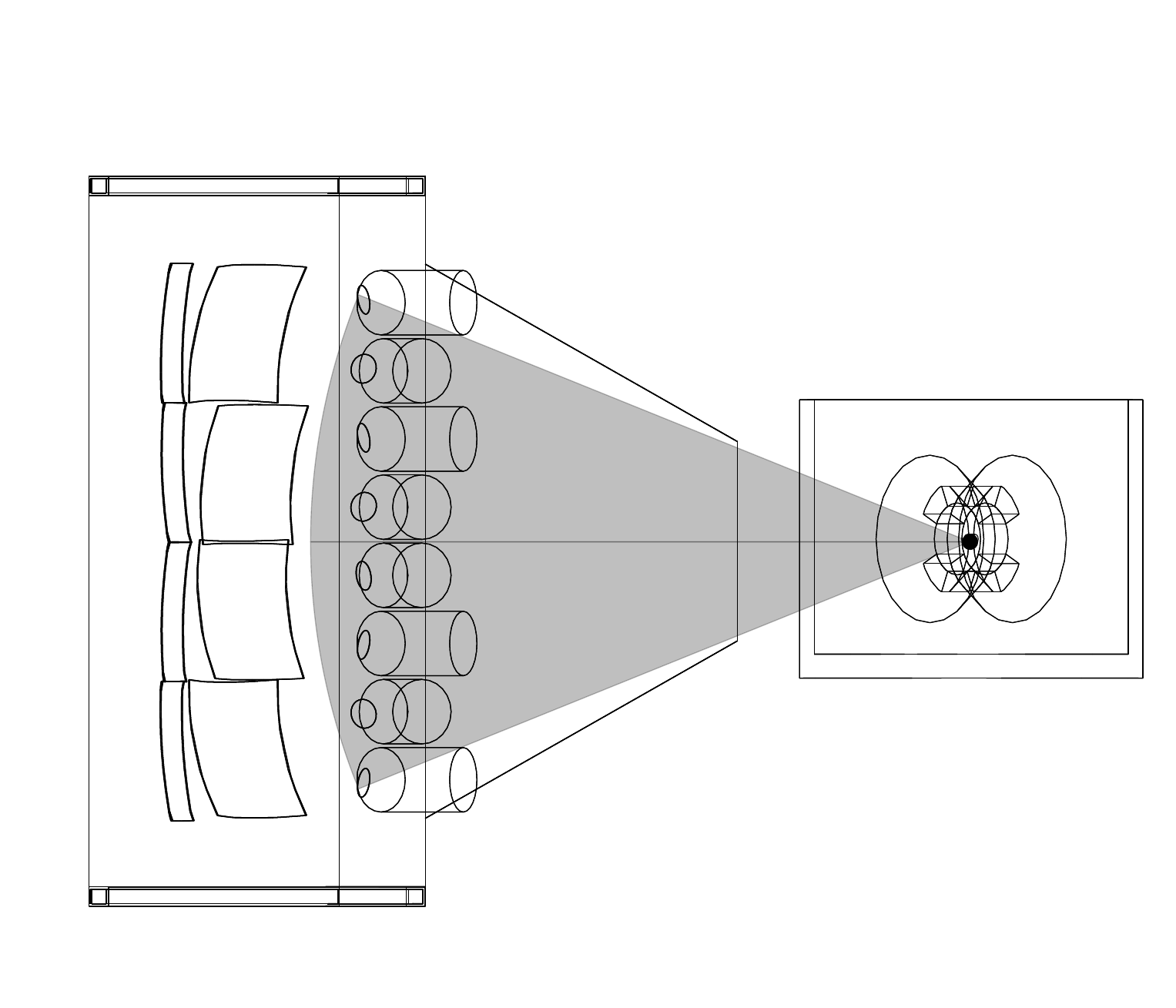} }
 \caption{\Cerenkov{} mirror alignment procedure.}
 \label{fig:mirroralign}
\end{figure}

Mirror alignment was completed before the detector was placed into its final 
position. A curved u-channel assembly (Figure~\ref{fig:mirrorHolder}) held the mirror 
and connected to the frame-mounted holding arms that adjusted the mirror positions.
The mirrors were aligned with the front window removed and using an 
attachment mounted to the very front of the detector, which held a small 
incandescent light bulb at the location of the target relative to the detector.  
Figures \ref{fig:mirroralign}a and \ref{fig:mirroralign}b show the alignment 
procedure using the front attachment which illuminated half of the detector at 
a time. A cap placed on the photo-multipliers protected them from light while 
the detector was open and during mirror alignment. These caps also had 
concentric circle targets on the outside to aid in the visual mirror alignment.  
With the attachment in place, the light bulb along with other tooling balls 
were used to survey the detector, allowing for very accurate final placement 
relative to the target as shown in Figure \ref{fig:mirroralign}c.

\begin{figure}[h]
\centering
\includegraphics[ trim = 40mm 65mm  5mm 41mm, clip,width=1\textwidth]{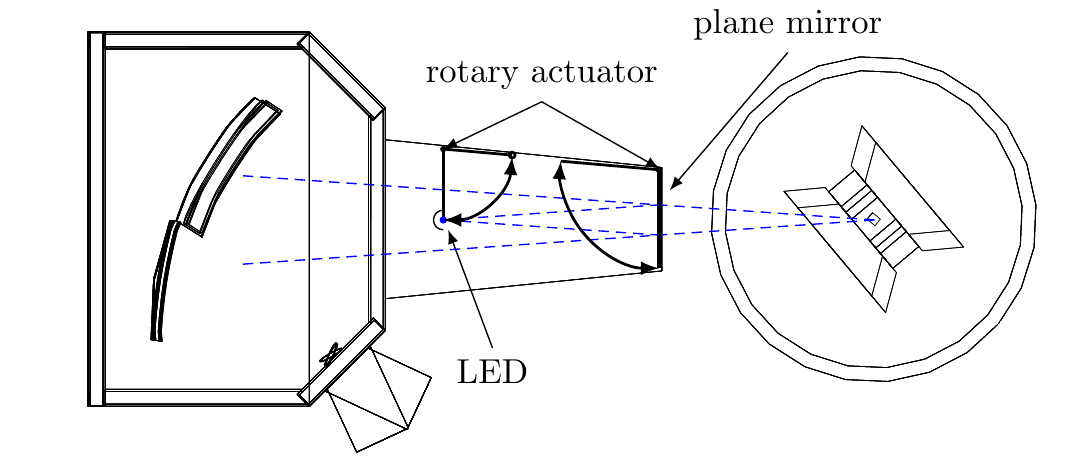}
\caption{Plan view of the \Cerenkov{} LED/mirror monitoring system.}
\label{fig:ledSystem}
\end{figure}

In order to calibrate and observe changes in the detector a LED/mirror system, 
shown in Figure~\ref{fig:ledSystem}, was installed inside the \Cerenkov{}.  An 
actuator rotated a thin mirror just inside the front window to reflect light 
from a LED, which also rotated to mirror the nominal target position.
When not in use the mirror and LED are rotated flush against the inside of the \Cerenkov{} snout.
This LED system was used to check for position shifts after the \Cerenkov{} was 
moved into its final position near the target.

\subsection{Photo-multipliers}
\label{sec:PMTs}

\begin{figure}[h]
 \centering
 \includegraphics[width=0.9\textwidth]{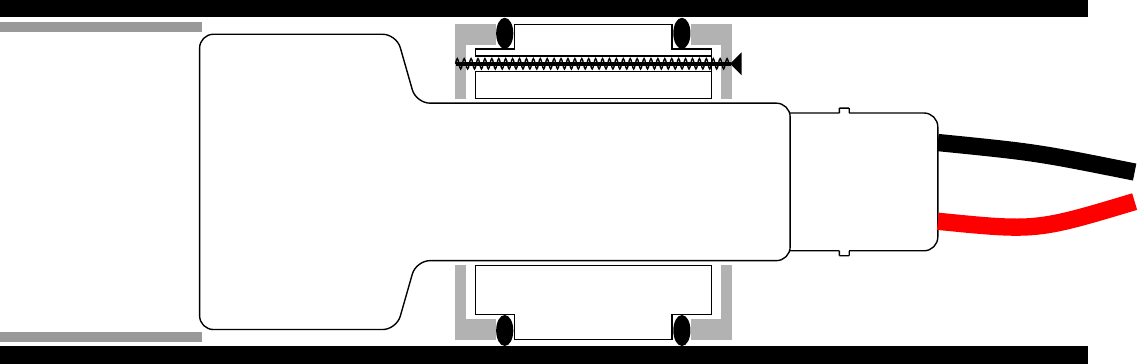}
 \caption{ A diagram (not to scale) of the $\mu$-metal PMT holder. \label{fig:pmtholder} }
\end{figure}

\begin{figure}[h]
 \centering
 \includegraphics[width=0.8\textwidth]{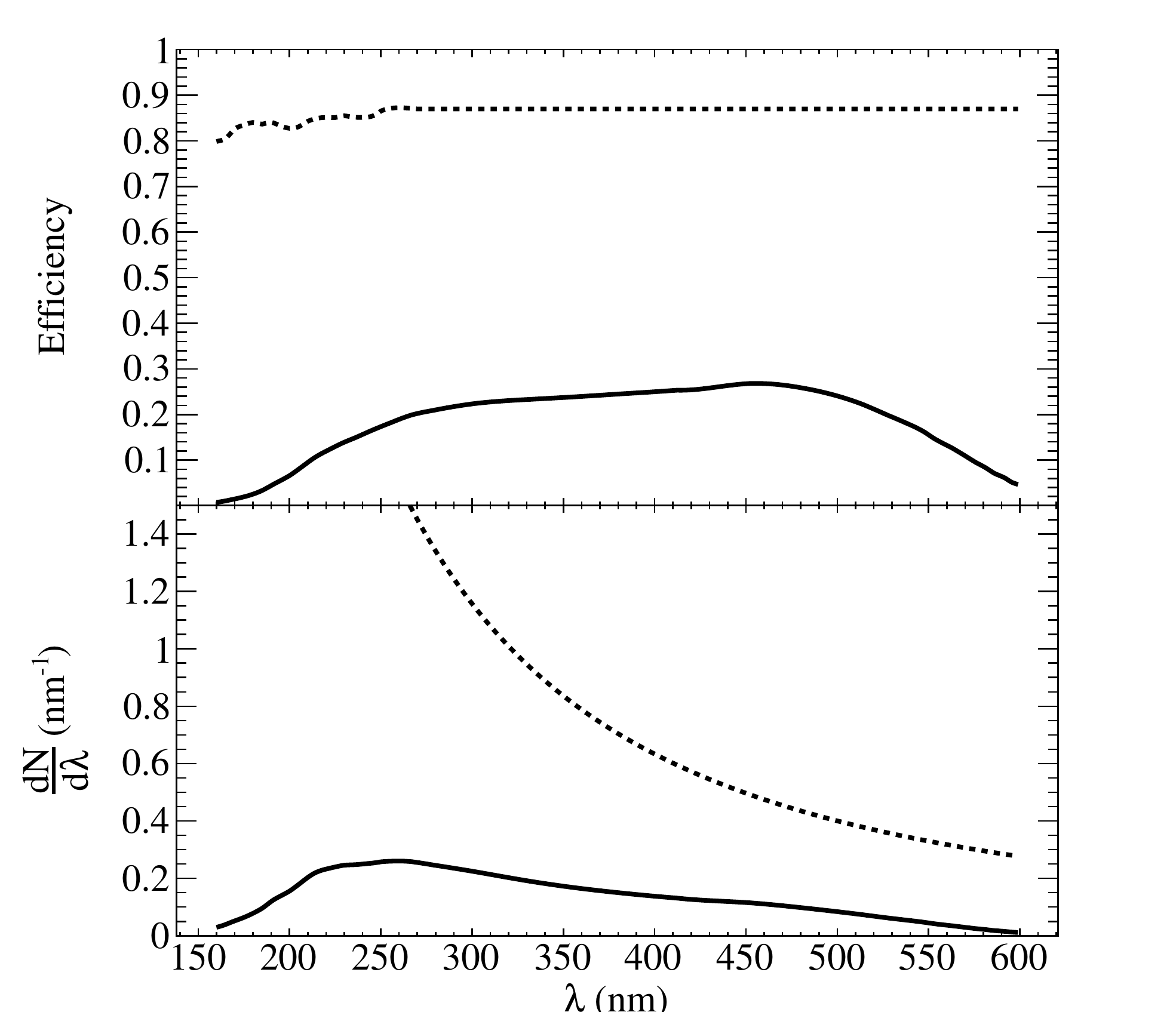}
 \caption{The mirror reflectivity (top, dashed), effective PMT quantum 
efficiency (top, solid line), the \Cerenkov{} spectrum (bottom, dashed), and 
the photo-electron yield integrand, Eq.~\ref{eq:cerSpectrum}  (bottom, solid 
line).   \label{fig:cerspectrum}}
\end{figure}

Three inch diameter Photonis XP4318B quartz window photomultiplier tubes were 
placed inside a $3$~mm thick mu-metal shield with its photocathode recessed 2 
inches from the end of the cylinder as shown in Figure~\ref{fig:pmtholder}.  
Quartz windows provided a complementary UV transparency needed to match the 
reflectivity of the mirrors.  Shown as the dashed-dotted curve in 
Figure~\ref{fig:cerspectrum}, the relative efficiency of the PMT is a 
combination of the photo-cathode quantum efficiency, the collection efficiency 
and quartz window transparency.  The \Cerenkov{} photon spectrum is given by 
the expression
\begin{equation}
 \frac{d^2N}{d\lambda dx} = 2 \pi \alpha (1-\frac{1}{ n^2 \beta^2})\frac{1}{\lambda^2}
\end{equation}
where $\alpha$ is the fine structure constant, $\beta$ is the velocity of the 
particle and $\lambda$ is the photon wavelength. The number of photons, N, is 
simply proportional to the gas length, L, which we take on average to be 
$1.3$~meters.

Neglecting the N$_2$ gas absorption, each scattered electron causes roughly 185 
photons to fall onto the mirrors and about 155 of these reflect towards the 
photo-multiplier tube. From such a light pulse about 20 photo-electrons should 
be counted.  This result can be calculated by integrating the solid curve in 
Figure~\ref{fig:cerspectrum} which is given by 
\begin{equation}\label{eq:cerSpectrum}
 \frac{dN}{d\lambda} = 2 \pi \alpha (1-\frac{1}{ n^2 
\beta^2})\frac{\mathrm{L}}{\lambda^2}  \eta(\lambda) R(\lambda)
\end{equation}
where $R$ is the reflectivity and $\eta$ is the relative efficiency of the PMT.

\subsection{Gas System and Tank Construction}
\label{sec:Gas_System} 

As previously mentioned, N$_2$ was selected as the \Cerenkov{} radiator gas due 
to the very high energy threshold for \Cerenkov{} light production from 
particles heavier than an electron.  Although air is mostly N$_2$ gas, it 
contains moisture and other gases (such as oxygen) which attenuate and distort 
propagating photons.

Using a dry N$_2$ gas source, a controller monitored and regulated the gas 
pressure inside the tank over the course of the experiment. A pressure 
transducer measured the pressure relative to the current atmospheric pressure 
and the controller opened and closed the appropriate solenoid valves to 
maintain a differential pressure of a few torr.  This slight constant 
overpressure reduced the possible contamination from the outside gases.

Based on initial measurements, the average delivery flow rate of gas while 
maintaining a differential pressure of 10 torr is less than 0.1 Standard Cubic 
Foot per Hour (SCFH). For an ideal hermetic and perfectly sealed \Cerenkov{} 
tank this rate would be zero, and therefore this rate is also referred to as a 
\textit{leak rate}. This is the rate that gas will enter the \Cerenkov{} tank 
when the manometer calls for more gas and automatically opens the fill solenoid 
valve to maintain a set point pressure.

\begin{figure}[h]
\centering
\includegraphics[width=1\textwidth]{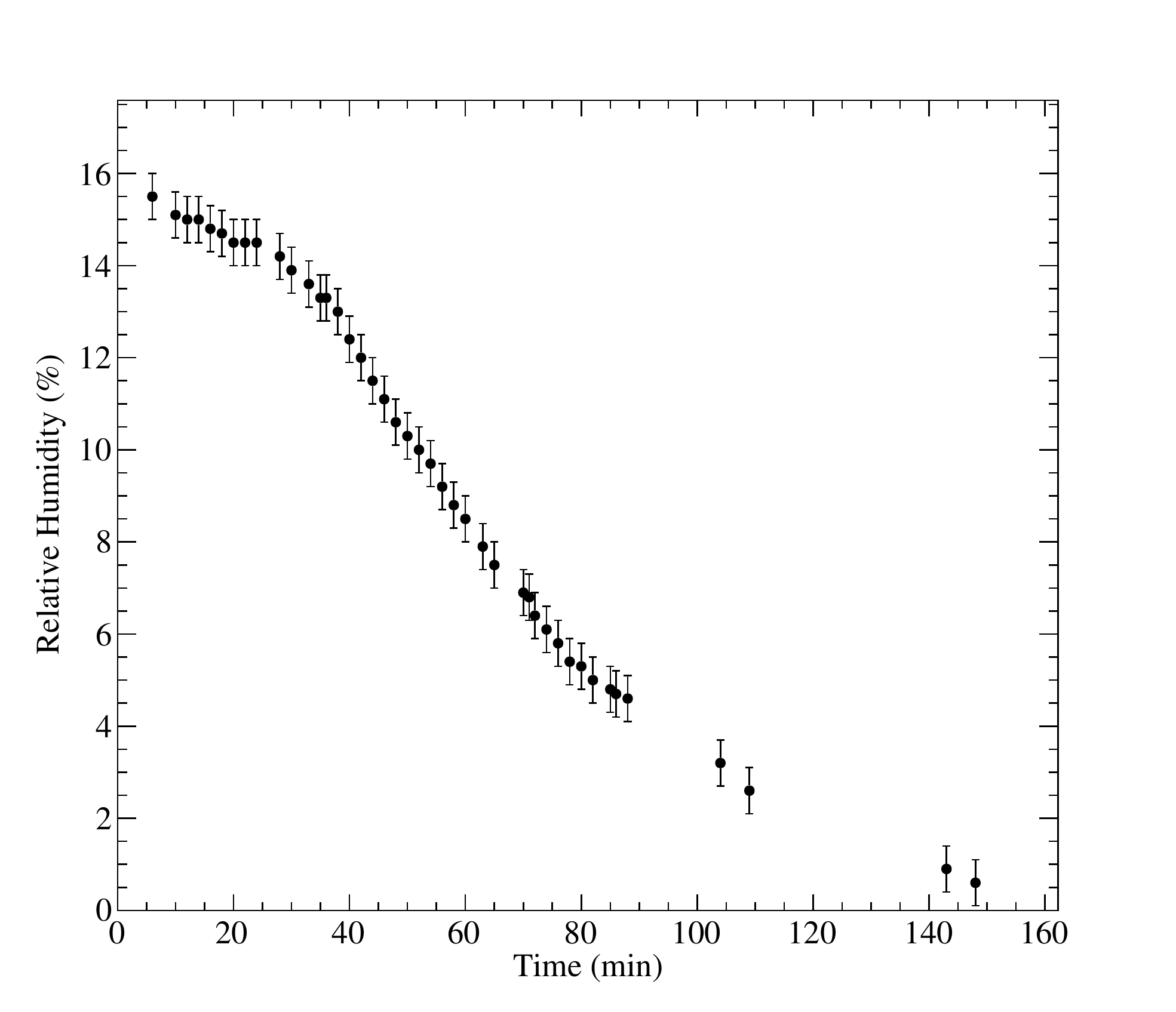}
\caption{Relative humidity while flushing at a differential pressure of about $10$~torr and a gas flow rate of $1.1$~SCFM.  \label{flushtest3}}
\end{figure}

Flushing the tank was important once it was sealed to purge all atmospheric gas 
present. The controller was placed in flush mode and the relative humidity of 
the venting gas monitored. A relative humidity of less than a few percent was  
achieved in about 2 hours as shown in figure~\ref{flushtest3}.
Desiccant was placed at the bottom of the detector to help remove any remaining moisture from the gas.

The tank's frame was constructed of welded and leak-checked two inch square 
aluminum tubing.  Attached to the front of the frame was the snout which 
narrows down to the front window frame holder. Attached to this frame holder 
was a thin piece of aluminum which is completely opaque, painted flat black, 
and framed by an aluminum bracket. The side panels of the main tank's frame 
were detachable, as was the rear window, a 1/16 inch aluminum sheet held by a 
large bracket.  Neoprene gaskets and O-rings were used throughout when 
attaching or sealing any panels or flanges. 

Two large lead walls were constructed to shield the PMTs from background.  The 
first sat between the target and \Cerenkov{} PMTs to shield them from 
scattering originating at the target. A second wall on the opposite side 
shielded the PMTs from secondary sources of scattering located around the beam 
dump.


\section{Computer Simulations}
\label{sec:simulation}

A GEANT4 \cite{Agostinelli:2002hh} simulation was developed to simulate the 
full BETA detector package called BETAG4\cite{BETAG4}. In addition to the BETA 
detectors, various aspects of the experiment and apparatus were implemented 
such as the target geometry, materials, magnetic field and emulation of the 
various triggers and scalers. Beyond providing insight into the behavior of the 
\Cerenkov{} counter alone, the simulation was used to explore various 
correlations with other detectors, observe the effects of extra material 
thicknesses, and study the event reconstruction from the calorimeter back to 
the target for each target polarization direction.

One important study undertaken using the simulation was to investigate the 
effect of the extra material in front of the \Cerenkov{} counter. This extra 
material thickness is almost entirely due to the forward tracker (which roughly 
totals to the same thickness in radiation lengths as the ammonia target).  
Beyond the small energy loss in this material, the primary problem comes from 
photons, which produce electron-positron pairs in the extra material. These 
pairs can produce a track that is misidentified as a scattered electron by the 
\Cerenkov{} counter.  To understand the effects of these background events they 
are classified by the location of their vertex and fall in two categories.

\begin{figure}[h]
\centering 
\includegraphics[trim = 4mm 4mm 4mm 14mm, clip,width=\textwidth]{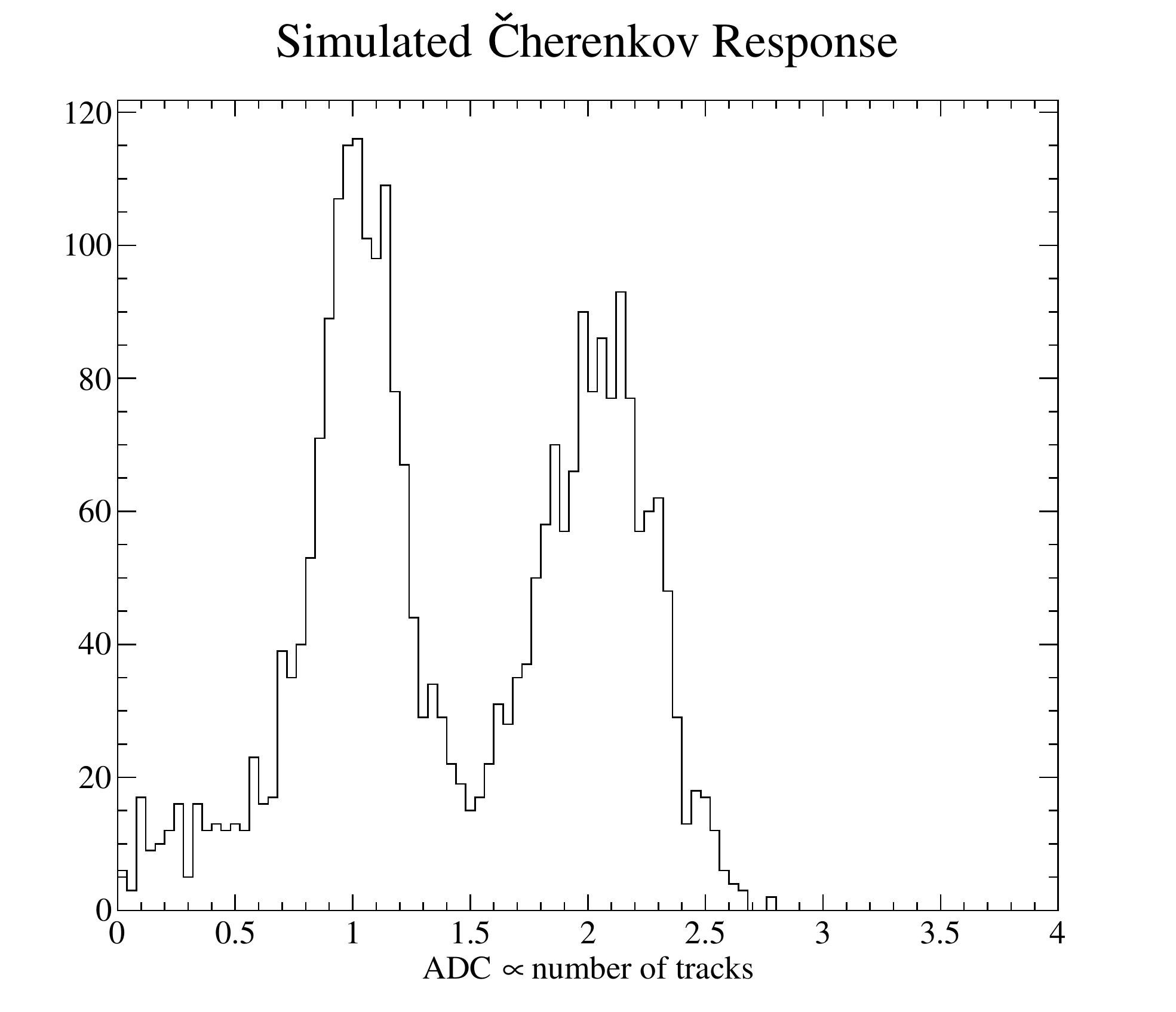}
\caption{Simulated \Cerenkov{} counter ADC spectrum for isotropic photons on 
BETA with energies uniformly sampled between $0.5$~GeV and $4$~GeV.  The ADC 
values are normalized by the average ADC single  for a single electron track.  
\emph{Intra-target} pairs produced within the target material in the central 
region of the target vacuum chamber are exposed to a very strong magnetic 
field. This field separates the electron and positron that at least one of the 
particles is ejected from BETA's acceptance. \emph{Intra-target} pairs dominate 
the first peak located around 1 track and \emph{extra-target} pairs are 
responsible for the second (double track) peak.
\label{fig:cerSimGamma}}
\end{figure}

Photons produce pairs within the target material in the central region of the 
target vacuum chamber, where the target magnetic field strength is quite large 
compared to the forward tracker location, just outside of the vacuum chamber.  
For these \emph{intra-target} pairs, the strong magnetic field significantly 
deflects (in opposite directions) the electron and positron trajectories, so 
much so, that even the high energy pairs almost never produce two tracks in 
BETA.  

However, \emph{extra-target} pairs produced in the forward tracker material 
generate nearly undeflected track into BETA, and more importantly, the pair  
produces \emph{twice} the amount of \Cerenkov{} photons as a good scattered 
electron track. This is clearly demonstrated in the simulation results of the 
ADC spectrum shown in Figure~\ref{fig:cerSimGamma}. We will return to the 
discussion of the pair symmetric background in 
section~\ref{sec:PairBackground}.


\section{Calibration and Commissioning}

The PMTs were calibrated to about 100 ADC channels per photo-electron. 
Following a technique of photo-electron counting described in 
\cite{Dossi:1998zn}, the charge response of each PMT was modeled as the sum of 
a pedestal Gaussian, single electron response function, 
$\text{SER}(\mathrm{x})$, and a multiple photoelectron response 
$M(\mathrm{x})$. The pedestal provides the so-called ``noise'' function which 
is convoluted with the single electron response.  To account for the non-ideal 
first dynode response and dynode noise, an exponential function is added to the 
ideal single (photo-) electron Gaussian function.  This ideal single electron 
response function is defined as \begin{equation}
   \text{SER}_0(\mathrm{x}) = 
\begin{cases}
   \displaystyle \frac{p_E}{A}~\mathrm{e}^{-\frac{\mathrm{x}-\mathrm{x}_p}{A}} 
   + \frac{1-p_E}{g_N 
   \sqrt{2\pi\sigma_0}}~\mathrm{e}^{-\left(\frac{\mathrm{x}-\mathrm{x}_0-\mathrm{x}_p}{\sqrt{2}\sigma_0}\right)^2 
   }
& \mathrm{x} > \mathrm{x}_p \\
       0 & \mathrm{x} < \mathrm{x}_p
\end{cases}
\end{equation}
where $\mathrm{x}_p$ is the pedestal peak position,
  $\sigma_p$ is the pedestal width,
  $p_E$ is the fraction of events which fall under the exponential,
  $A$ is the decay constant of exponential,
  $\mathrm{x}_0$ is the single photo-electron peak position, and 
  $\sigma_0$ is the width of the single photo-electron peak.

\begin{figure}[h] 
 \centering
 \includegraphics[width=\textwidth]{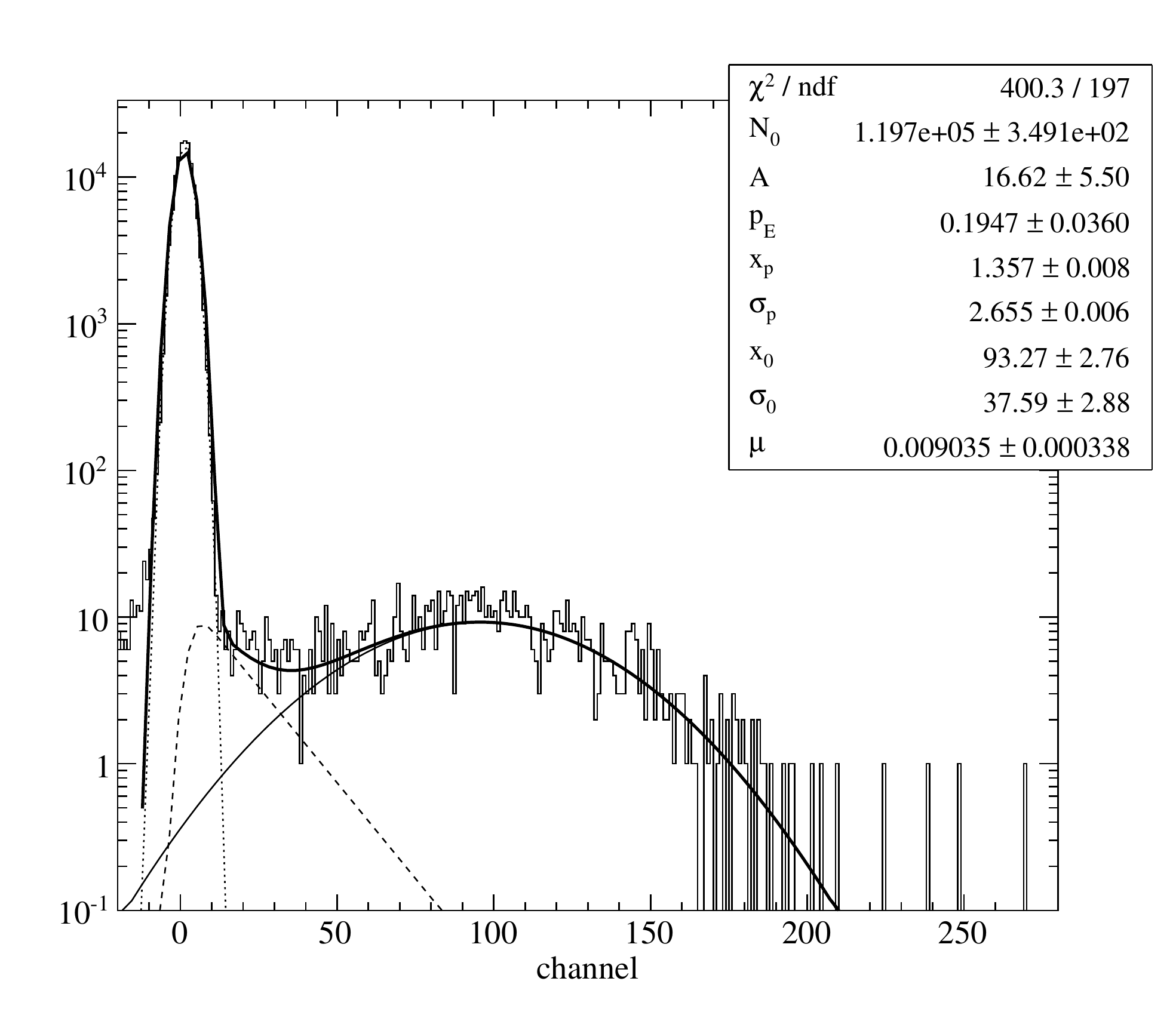}
 \caption{Single photo-electron response fit result (solid) using 
equation~\ref{eq:PMTfit}, neglecting the $M(\mathrm{x})$ term. Also shown are 
the contributions from the pedestal (dotted), exponential (dashed), and Gaussian 
(solid) part of $SER(\mathrm{x})$. \label{fig:onePEsingle}}
\end{figure}

In order to take into account non-ideal issues such as electrical noise or ADC resolution, a noise function, defined as 
\begin{equation} \label{eq:PMTNoise}
\displaystyle  \text{Noise}(\mathrm{x}) = \frac{1}{\sqrt{2\pi}\sigma_p} 
\mathrm{e}^{-\left(\frac{\mathrm{x}-\mathrm{x}_p}{\sqrt{2}\sigma_p}\right)^2},
\end{equation}
is convoluted with the ideal single electron response to yield a realistic 
single (photo-)electron response function.
\begin{equation} \label{eq:PMTSER}
 \displaystyle\text{SER}(\mathrm{x}) =  (\text{Noise}\otimes\text{SER}_0)(\mathrm{x})
\end{equation}
This function used in the model of the full ADC spectrum,   
\begin{equation} \label{eq:PMTfit}
\displaystyle f(\mathrm{x}) = N_0( P(0) \text{Noise}(\mathrm{x}) + P(1) \text{SER}(\mathrm{x}) + M(\mathrm{x}))
\end{equation}
where $P(0)$ and $P(1)$ are the probabilities for 0 and 1 photo-electron response, and  $N_0$ is a normalization. 
The multiple photo-electron response, $M(\mathrm{x})$ is
\begin{equation} \label{eq:Mx}
\displaystyle   M(\mathrm{x}) = \sum_{n=2}^{N_{M}} 
\frac{P(n;\mu)}{\sqrt{2n\pi}\sigma_1} \mathrm{e}^{-\frac{(\mathrm{x} - n 
      \mathrm{x}_1 
- \mathrm{x}_p)}{2 n \sigma_1^2}}
\end{equation}
where 
  $\mu$ is the average number of photo-electrons,
  $\mathrm{x}_1$ ($\simeq\mathrm{x}_0$) is the average number of channels per photo-electron,
  $\sigma_1$ is the average width of the (Gaussian) PMT response to a single photo-electron,
  $N_{M}$ is the cut off in the sum which should be much greater than $\mu$, 
and
  $P(n;\mu)$ is the Poisson probability distribution for $n$ photo-electrons with mean value $\mu$.

In order to determine the location of the single photo-electron peak, the LED 
pulser was set very low, $\mu < 0.01$. The pedestal amplitude was roughly 
determined by counting the number of events within the pedestal, $N_p$, and 
then taking the ratio to yield the zero photo-electron probability 
$P(0)=N_p/N_{Tot}$, where $N_{Tot}$ is the total number of LED triggers.

Since $\mu$ is much less than unity ($\mu \ll 1$), the contribution of multiple 
photo-electrons in Eq.~\ref{eq:PMTfit} can be neglected when fitting the LED 
data. An example of such a fit is shown in figure~\ref{fig:onePEsingle}. 
Fitting yields values for the parameters  $\mathrm{x}_p$, $\sigma_p$, 
$\mathrm{x}_0$, and $\sigma_0$ for each PMT.

\begin{figure}[h]
 \centering
 \includegraphics[width=\textwidth]{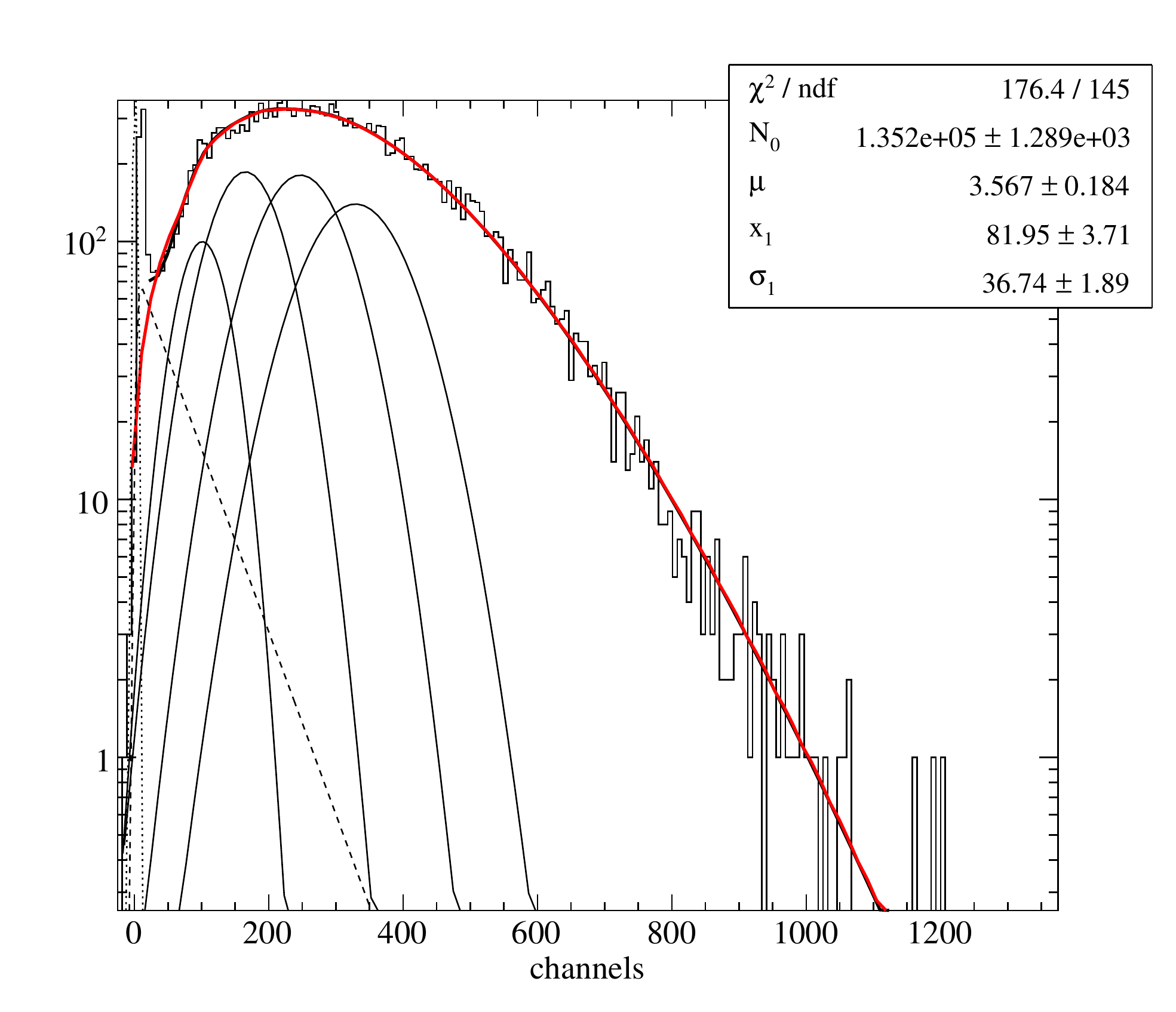}
 \caption{A fit to the photomultiplier ADC spectrum with an LED amplitude of a 
 few photo-electrons. It shows the various terms including the pedestal 
 (dotted), exponential (dashed), single photo-electron peak (solid), and the 
 first few terms in the multiple photo-electron response (solid).  
 \label{fig:LEDfit}}
\end{figure}

Next, in order to calculate the contribution from multiple photoelectrons the 
parameters in equation~\ref{eq:Mx} are needed. To first approximation 
$\mathrm{x}_1 \simeq \mathrm{x}_0$ and $\sigma_1 \simeq \sigma_0$, however, for 
larger signals this approximation may not hold. In order to check this, the LED 
pulser amplitude is turned up to $\mu\simeq5$. Subsequently using the fixed parameters 
$\mathrm{x}_p$, $\sigma_p$, $\mathrm{x}_0$, and $\sigma_0$, a spectrum fit is 
performed.
From this fit the values of $\mu_{1}$ and $\sigma_{1}$ are determined as shown in Figure~\ref{fig:LEDfit}.

\begin{figure}[h]
 \centering
\includegraphics[trim = 0mm 0mm 0mm 10mm,clip,width=0.8\textwidth]{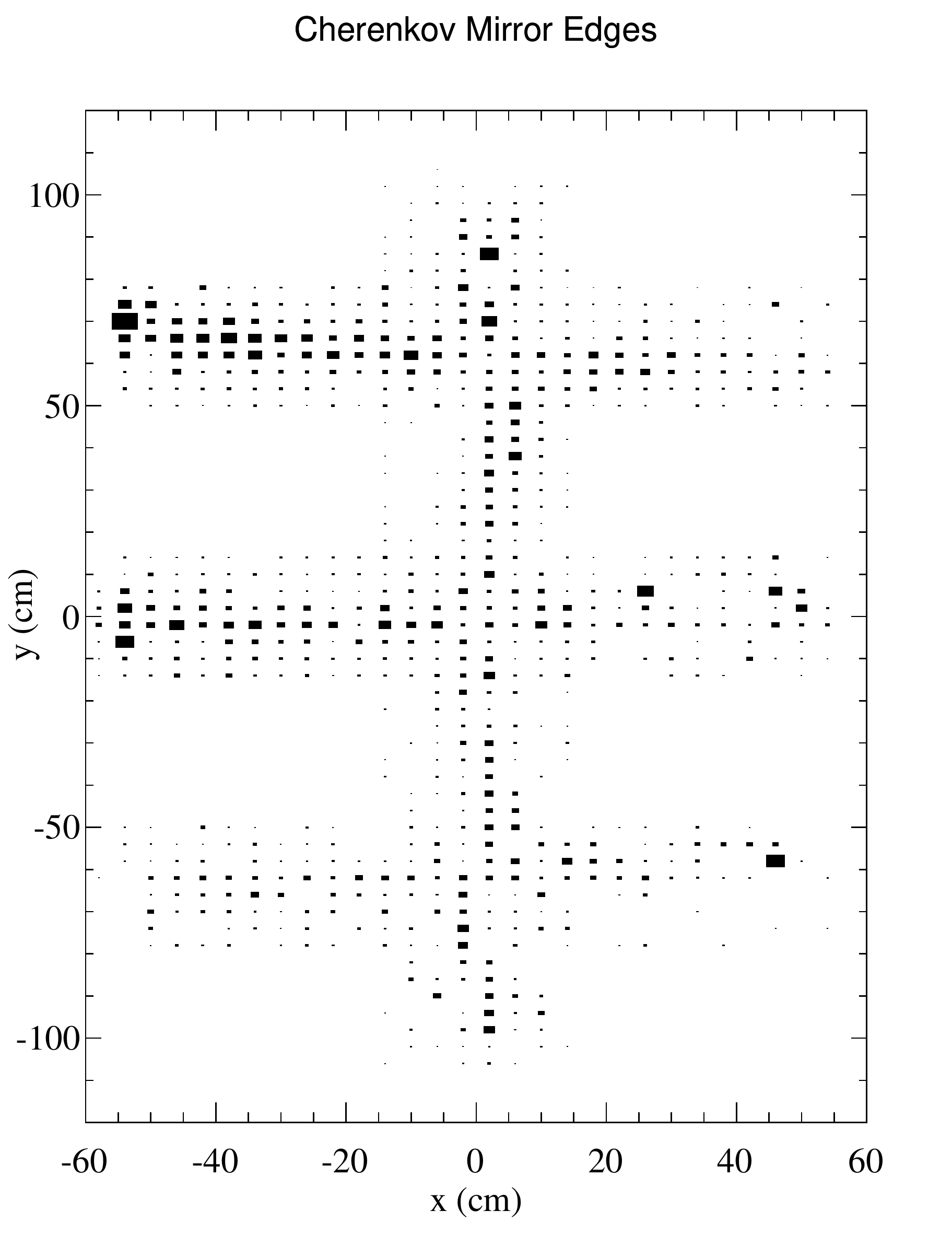}
 \caption{The mirror edges are shown in the calorimeter cluster positions when 
a TDC hit is required on two adjacent mirrors for a cluster. Also note how the 
size of the \Cerenkov{} cone can be estimated from the distribution of events 
across the mirror edges.\label{fig:edges}}
\end{figure}

For some events the \Cerenkov{} cone is split between multiple mirrors, thus 
locating the mirror edges is important for knowing how to form an appropriate 
ADC sum.  Plotting the calorimeter's cluster positions while requiring a TDC 
for at least two a adjacent mirrors produces a projection of the mirror edges 
as shown in Figure~\ref{fig:edges}.  When a cluster falls within the region of an 
edge, the corresponding mirror's ADCs are added to form the \Cerenkov{} sum 
associated with the cluster.


\section{Performance }
\label{sec:Performance}

\subsection{Data Acquisition and Analysis}
\label{sec:DAQ}

The eight analog \Cerenkov{} signals were summed using a Lecroy 428F Quad 
Linear Fan-In/Fan-Out NIM module.  Using a single discriminator, the analog 
sum, in coincidence with the calorimeter, formed the primary trigger used 
during the SANE experiment. The performance of the SANE Gas \Cerenkov{} is 
reported for two configurations which are characterized by the polarized 
target's magnetic field orientation. 

The SANE experiment used a 5.1 Tesla superconducting magnet with its field 
aligned  anti-parallel or transverse ($80^{\circ}$) to the beam direction. The 
\Cerenkov{} detector was exposed to very different fringe fields and background 
rates (Table~\ref{tab:cerRates}) with each target field configuration. This 
difference in rates is primarily due to lower energy forward scattered 
particles circling along the polarization direction of the magnet, which in the 
parallel case travels along the beam direction towards the beam dump. However, 
for the transverse polarization, these low energy particles are directed 
towards the detector thus increasing the background compared to the parallel 
orientation.  The individual PMT rates were roughly proportional to the solid 
angle covered by the corresponding mirror.

\begin{table}[h]
\begin{center}
\begin{tabular}{|l|r|r|}
\hline
Channel & Parallel   & Perpendicular \\ 
\hline
  1 & 33  kHz & 1532  kHz \\
  2 & 72 kHz & 685 kHz\\
  3 & 41 kHz & 960 kHz\\
  4 & 74 kHz & 473 kHz\\
  5 & 43  kHz& 587 kHz\\
  6 & 86 kHz & 516 kHz\\
  7 & 53 kHz & 1023 kHz\\
  8 & 60 kHz & 338 kHz\\
\hline
 Lower 4 sum & 103 kHz&  928 kHz\\
 Middle 4 sum & 131 kHz & 1059 kHz\\
 Upper 4 sum & 114 kHz & 1125 kHz \\
\hline
 Total sum & 206 kHz & 1795 kHz \\
\hline
\end{tabular}
\caption{Rates during normal operations for the two configurations. \label{tab:cerrates}}
\label{tab:cerRates}
\end{center}
\end{table}

The detector commissioning occurred with the target polarization in the 
transverse orientation due to target magnet problems. Therefore, we begin by 
reporting the performance during the first part of the experiment.


\subsection{Transverse Field Orientation}
\label{sec:transverse}

The first \Cerenkov{} counter data were taken with the target in transverse 
field orientation.  A few unanticipated problems arose during initial 
commissioning of the detector.  

The spherical mirror PMTs had a slightly larger than anticipated longitudinal 
magnetic field component which degraded the efficiency of the PMTs on that 
side.  Their performance degradation was monitored using the LED pulsing 
system.  To help mitigate this inefficiency, a large 3/4-inch iron plate was 
quickly mounted on the front of the lead shielding wall to act as a magnetic 
yoke.  The performance improved markedly after the plate was installed.

\begin{figure}[h]
\centering
\includegraphics[width=0.9\textwidth]{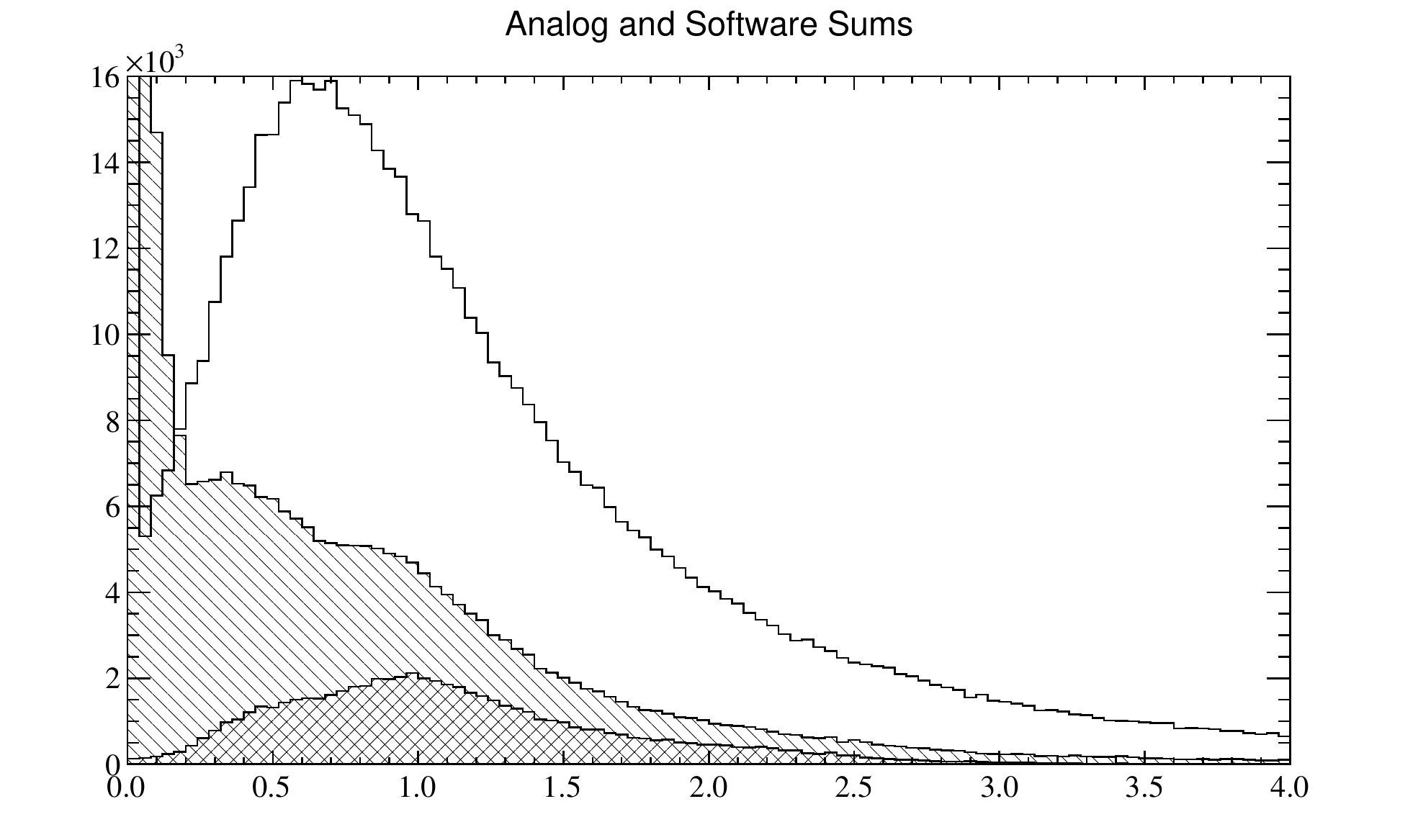}
\caption{ADC spectra during transverse field configuration are shown: the 
analog sum of all eight mirrors (no hatches), cluster correlated ADC sum 
(single hatches), and cluster correlated ADC sum with a \Cerenkov{} TDC cut.
\label{fig:PerpADCSums}}
\end{figure}

The transverse field production runs are  characterized by high background 
rates on all detectors, including the \Cerenkov{} counter. Therefore, clean 
event selection required a correlated calorimeter cluster and hodoscope hit.
The necessity of these cuts can be seen in Figure~\ref{fig:PerpADCSums}.  
Comparing the analog sum, which was used in the trigger, to the software sum  
of the cluster correlated mirrors, there remained a substantial background.  
With the addition of a \Cerenkov{} timing cut, the background is dramatically 
reduced, indicating a background composed of uncorrelated, low energy electrons 
or positrons, consistent with the description of nearly trapped particles in 
the target's transverse magnetic field.

\begin{figure}[h]
 \centering
 \includegraphics[width=0.8\textwidth]{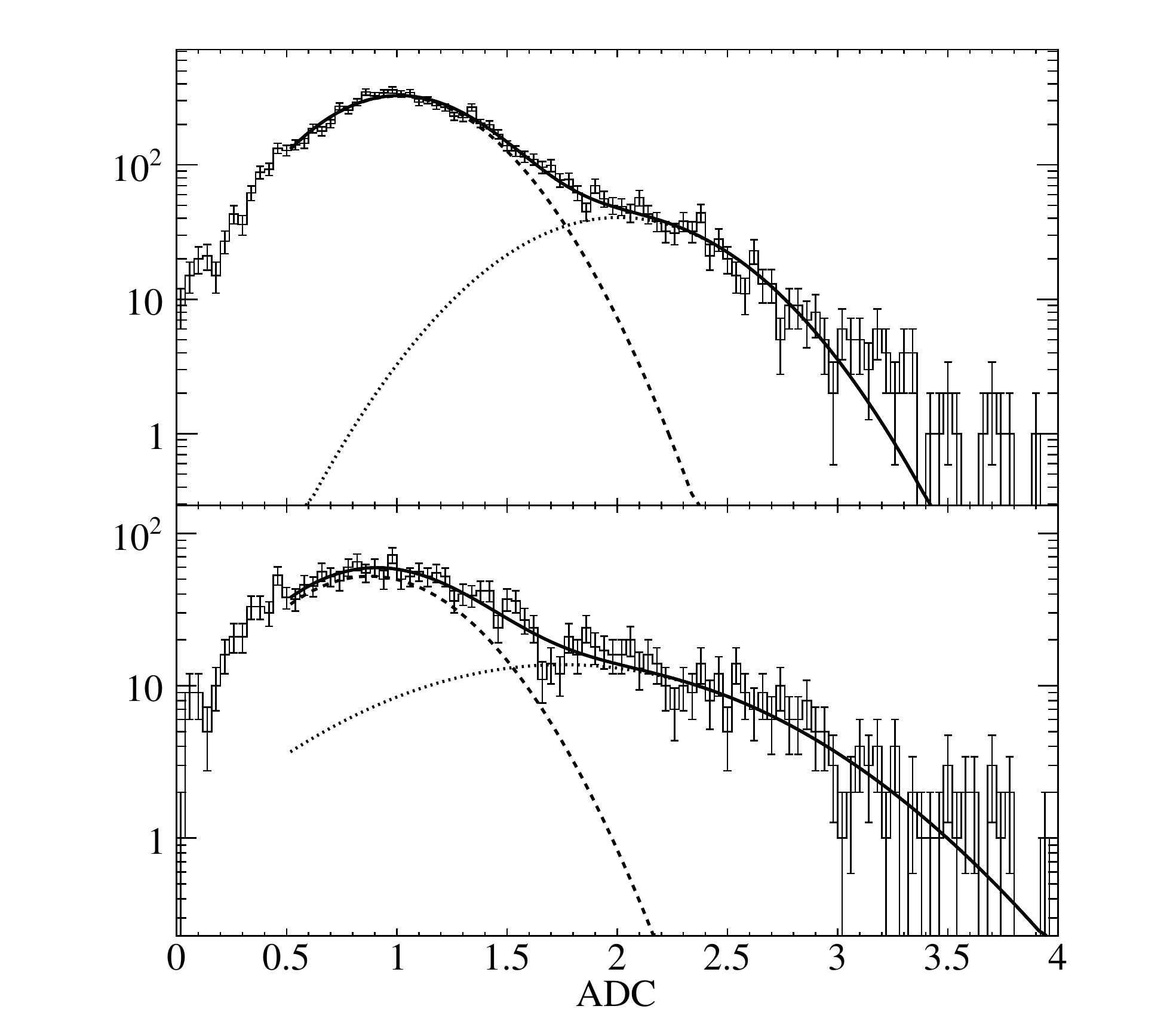}
 \caption{\Cerenkov{} counter ADC spectrum from a transverse field 
configuration run for all the toroidal mirrors (top) and spherical mirrors 
(bottom).\label{fig:PerpClean}}
\end{figure}


\subsection{Anti-parallel Field Orientation}

\begin{figure}[h]
\centering
\includegraphics[width=0.8\textwidth]{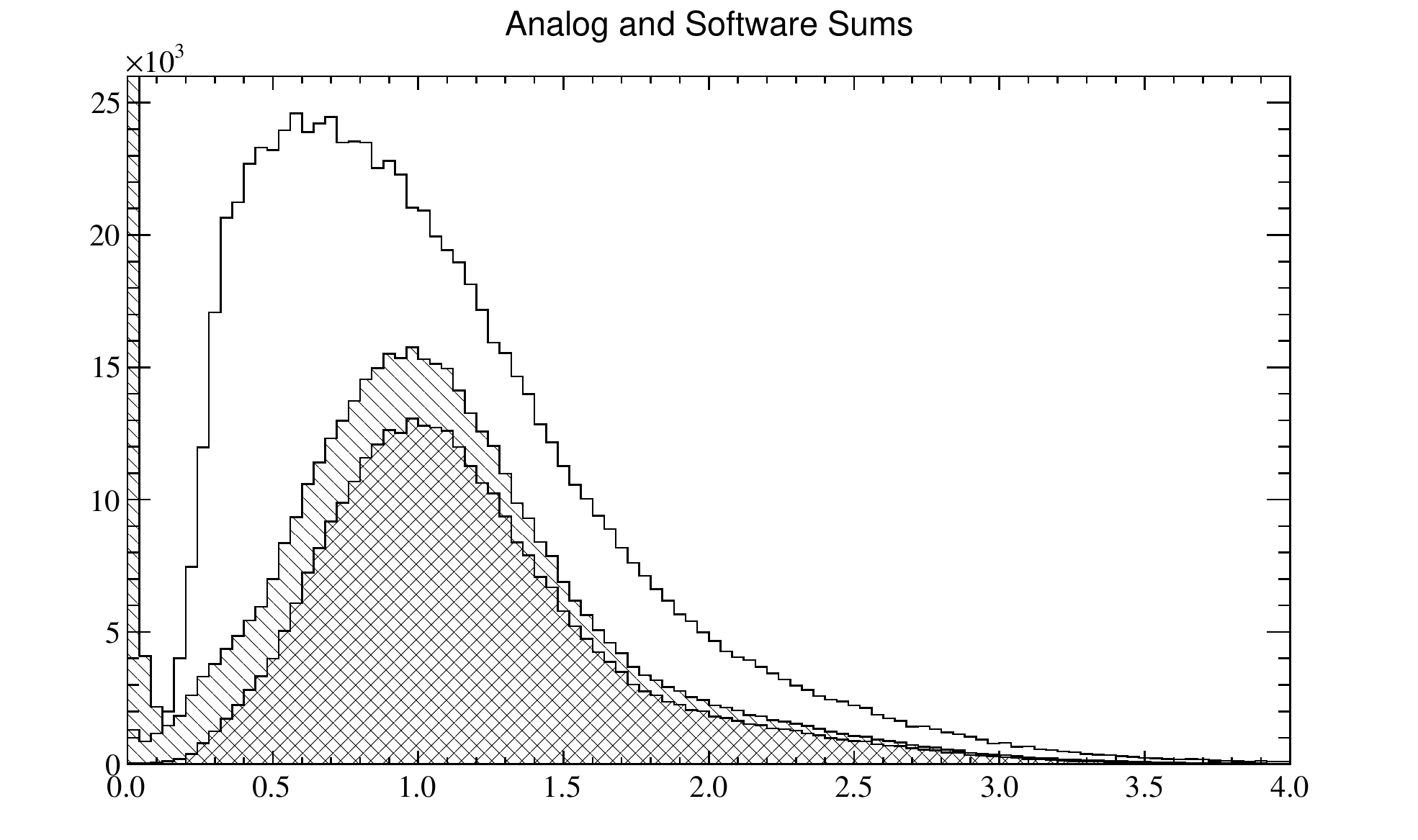}
\caption{ The histograms above show the calibrated ADC spectra for an 
anti-parallel configuration. The ADC spectrum of the analog sum of all eight 
mirrors (no hatches), cluster correlated ADC sum (single hatches), and cluster 
correlated ADC sum with a \Cerenkov{} TDC cut. \label{fig:ParaADCSums}}
\end{figure}

The background rates were significantly lower during parallel target running.  
Unlike the transverse configuration the software sums have much less 
background. Even without the aid of BETA detector event selection, the 
\Cerenkov{} spectrum appeared relatively free of background as shown in 
Figure~\ref{fig:ParaADCSums}.

\begin{figure}[h]
\centering
\includegraphics[width=1.0\textwidth]{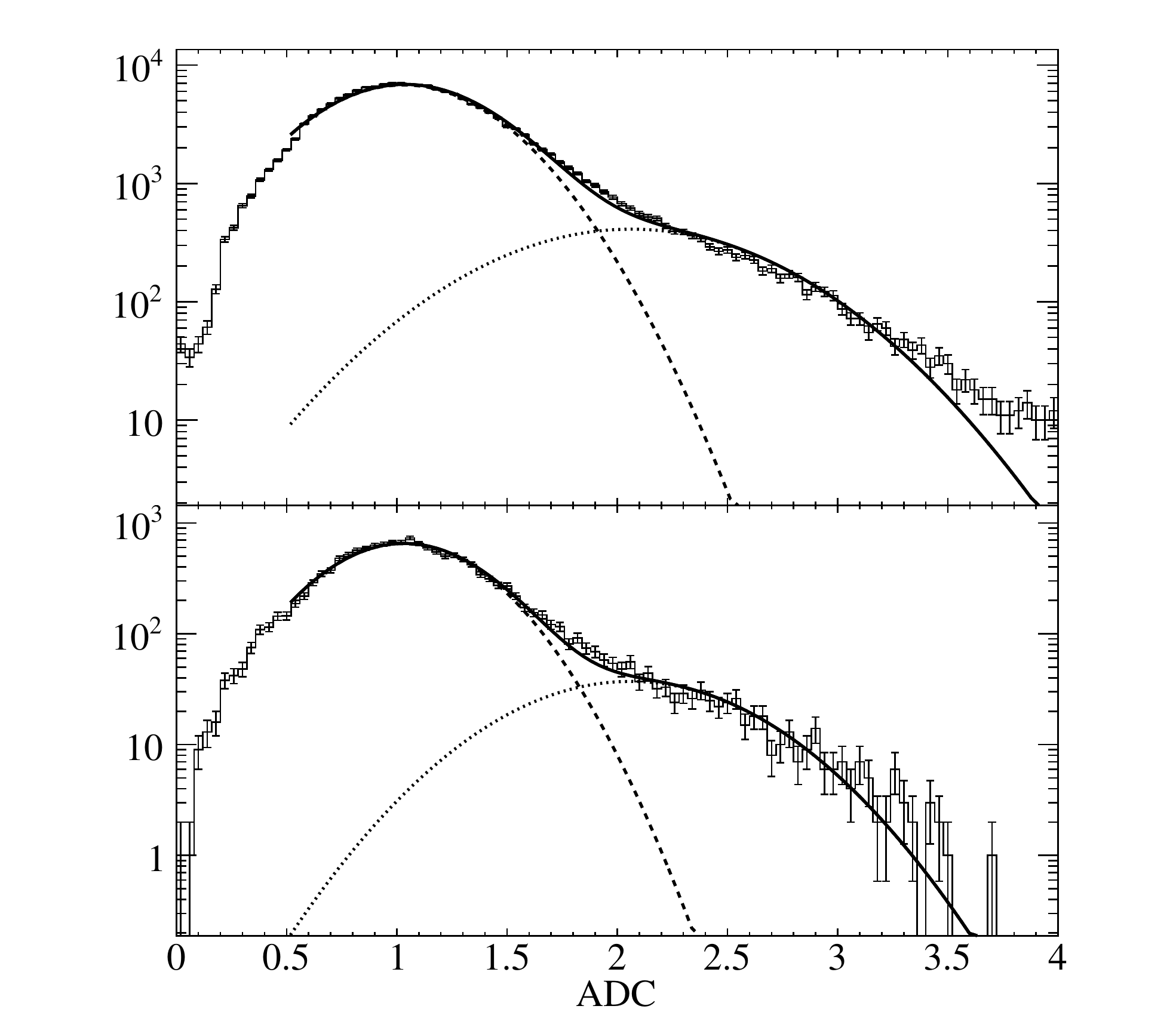}
 \caption{Same as Figure~\ref{fig:PerpClean} for a anti-parallel field 
 run.\label{fig:ParaClean}  }
\end{figure}


\subsection{Pair Symmetric Background Identification}
\label{sec:PairBackground}

\begin{figure}[h]
 \centering
 \includegraphics[width=\textwidth]{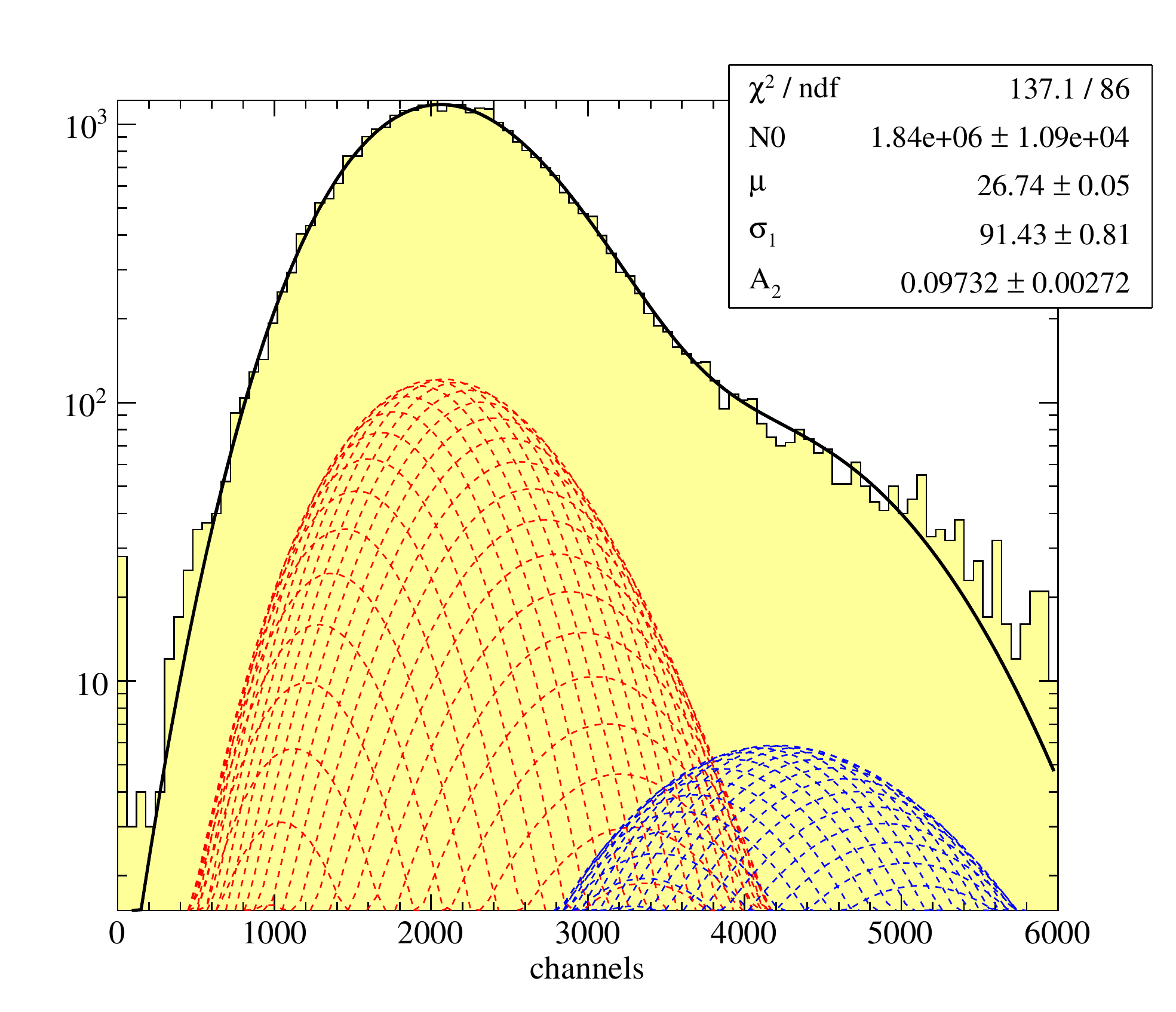}
 \caption{A typical ADC spectrum showing the one and two track peaks and the 
convolution fit.  \label{fig:multiFit6}}
\end{figure}

It can be seen from the ADC spectra of Figures~\ref{fig:PerpClean} and 
\ref{fig:multiFit6} that there was a significant contribution from tracks that 
have twice the amount of light than a single electron track produces.
Most background events come from electron positron pair-production which can 
originate from one of two locations as previously discussed 
(Section~\ref{sec:simulation}).

Simulation has shown that \emph{intra-target} pairs produced at the target are 
very unlikely to produce a double hit due to the strong (and opposite) 
deflections through the target magnetic field. This background cannot be 
removed and must be accounted for in the data analysis. 

Only \emph{extra-target} pairs remain close enough to each other to appear as a 
single cluster at the calorimeter. Fortunately, placing an ADC window cut on 
the \Cerenkov{} ADC spectrum alone removes a significant amount of these pair 
events and allows for an accurate estimation of the cuts efficiency. The ADC 
window cut consists of the usual lower ADC limit and an upper limit located 
between the single and double track peaks. The precise location of the upper 
limit is a compromise between statistics and (extra-target) pair background 
rejection. For a systematic error limited measurement, pushing this limit 
closer to the single track eliminates a large fraction of the background, and 
by fitting the ADC spectrum as shown in Figure~\ref{fig:multiFit6}, the 
background contamination can be calculated by integrating the double track 
peak.


\section{Conclusion }
\label{sec:conclusion}

The SANE gas \Cerenkov{} detector operated successfully in a high luminosity 
environment with an open configuration and covered a large solid angle and 
momentum acceptance.  With a large photo-electron yield, it identified 
electrons with high efficiency. This large photo-electron yield provided enough 
ADC separation of double track events, allowing for novel event selection with 
the removal of (extra-target) pair symmetric background events. Although the 
\Cerenkov{} counter operated efficiently under both target configurations, the 
transverse field orientation proved most challenging due to higher background 
rates.

\section{Acknowledgements}
This work is supported by DOE grant DE-FG02-94ER4084

%
\bibliography{cherenkov} 




\end{document}